\documentstyle[11pt,aaspp4,psfig]{article}

\received{}
\accepted{}
\begin{document}
\def\kms{km~s$^{-1}$}
\def\cm2{cm$^{-2}$}
\def\Lya{Ly$\alpha$~}
\def\lya{Ly$\alpha$}
\def\Lyb{Ly$\beta$~}
\def\lyb{Ly$\beta$}
\def\Lyg{Ly$\gamma$~}
\def\Lyd{Ly$\delta$~}
\def\Lye{Ly$\epsilon$~}
\def\d{$d_5$~}
\def\Ly{Lyman~}
\def\ang{\AA~}
\def\gq{$\geq$~}
\def\zem{$z_{em}$~}
\def\zabs{$z_{abs}$~}
\def\etal{{\it et al.}~}

%%%%%%%%%%%%%%%%%%%%%%%%%%%%%%%%%%%%
%%%     Atomic Defs    %%%%%%%%%%%%%
%%%%%%%%%%%%%%%%%%%%%%%%%%%%%%%%%%%%
\def \HI   {\ion{H}{1}}
\def \DI   {\ion{D}{1}}
\def \HeII {\ion{He}{2}}
\def \CIV  {\ion{C}{4}}
\def \MgII  {\ion{Mg}{2}}
\def \FeII  {\ion{Fe}{2}}
\def \SiIII  {\ion{Si}{3}}

% -- 1. Title page !
%
\title{The Deuterium Abundance at z=0.701 towards QSO 1718+4807\altaffilmark{1}}

\author{David Tytler\altaffilmark{2}, Scott Burles\altaffilmark{3}, 
Limin Lu, Xiao-Ming Fan, 
Arthur Wolfe}
\affil{Department of Physics, and Center for Astrophysics and Space
Sciences \\
University of California, San
Diego \\
C0424, La Jolla, CA 92093-0424}
 
\author{Blair D. Savage}
\affil{Washburn Observatory \\ University of Wisconsin \\ 
475 N. Charter Street \\
Madison, WI 53706}

\altaffiltext{1}{Based on observations obtained with the NASA/ESA Hubble
Space Telescope
obtained by the Space Telescope Science Institute, which
is operated by AURA, Inc., under NASA contract NAS5-26555.}

\altaffiltext{2}{Visiting Astronomer, W. M. Keck Telescope, California
Association for Research in Astronomy}
 
\altaffiltext{3}{Present address: University of Chicago, Dept. of Astronomy \& Astrophysics, 5640 S. Ellis Ave., Chicago, IL 60637}
 
%
% -- 2.Abstract page!
%
 
\begin{abstract}
We present constraints on the deuterium
to hydrogen ratio (D/H) in the metal-poor
gas cloud at redshift $z=0.701$ towards QSO 1718+4807.
We use new Keck spectra in addition to Hubble Space Telescope (HST) and
International Ultraviolet Explorer (IUE) spectra. 
We use an improved redshift and a lower \HI~column density 
to model the absorption.
The HST spectrum shows
an asymmetric Lyman-$\alpha$ (\lya) feature which is produced by
either \HI~at a second velocity, or a high abundance of D.
Three models with a single simple H+D component give 
$8 \times 10^{-5} < \rm{D/H} < 57 \times 10^{-5}$ (95\%), 
a much larger range than reported by Webb \etal (1997a,b).  
A more sophisticated velocity distribution, or a second
component is necessary for lower D/H.
With two components, which could be a part of one absorbing structure,
or separate clouds in a galaxy halo, we find $\rm{D/H} < 50 \times 10^{-5}$. 
We do not know if this second component is present, but
it is reasonable because
40 -- 100\% of absorption systems with similar redshifts and \HI~column 
densities have more than one component.
\end{abstract}


\section{INTRODUCTION}

The absorption system at $z=0.701$ towards Q1718+4807 is well suited
for a determination of D/H from analysis of the line profiles of the
Lyman series.  We selected this line of sight to measure D/H for four
reasons:
(1) Low redshift Lyman lines are likely to be less contaminated 
by unrelated \HI~\Lya lines.
(2) The background QSO is bright and affords the possibility of 
high resolution UV spectroscopy with HST.
(3) IUE spectra of the QSO (\cite{lan93}) 
show a partial Lyman limit which is very steep,
which implies that the Lyman series lines are unusually narrow.
At high redshift most Lyman lines are broad enough that they 
swallow the D lines.
(4) Metal lines in Keck optical spectra were very weak. 
We now estimate that HST+FOS spectra limit the system's
metallicity to less than 1/100 solar. 

Webb \etal (1997a, 1997b, hereafter WCLFLV) 
have studied this system in detail and 
find D/H = 20 $\pm \, 5 \times 10^{-5}$. 
They assume the absorber is simple, with a single velocity component 
(e.g. one cloud in a halo) and they assume that it produces
lines with Voigt profiles. These assumptions are
justified by the simple appearance of the
\SiIII~and \MgII~lines and the steep Lyman Limit,
but they are ruled out by the asymmetric \Lya line, unless D/H is very
high.

Levshakov \etal (1998) use the same data as WCLFLV
and apply more sophisticated models which allow for correlations in
the velocity field (non-Voigt profiles).  
They use realistic parameters for the gas:
temperatures from $1.4 - 1.8 \times 10^4$ K, and turbulent velocities
of 18 -- 40 \kms.
Their models produce
asymmetric profiles for the \HI~\Lya absorption, with more \HI~absorption
and less D on the blue side of the \Lya line, and they find a lower 
estimate of D/H $\simeq 4.4 \times 10^{-5}$.  
These models allow only one velocity component, 
and  they assume that both the temperature and RMS turbulent velocity
are constant along the line of sight through this absorber.

In this letter, we 
present new Keck spectra, and analyze the range of D/H allowed by the
HST+GHRS, HST+FOS, and IUE spectra for this absorber.
In Section 2, we describe the spectroscopic observations and results 
from the four different instruments.
In Section 3, we investigate a set of four models to analyze the systematic
differences in the derived D/H between models.  We also assess the 
magnitudes of other possible systematic effects.

\section{OBSERVATIONS \& ANALYSIS} 

On March 3, 1995, the observations of QSO 1718+4807 (zem = 1.084, V = 15.3)
were obtained with the GHRS on HST.  Fifty-six five minute exposures of QSO
1718+4807, each covering 2050--2099 \AA~were taken with the G270M grating.
We used the small science aperture (SSA) to reduce the
blurring effects of the HST spherical aberration in these post-COSTAR
observations. COSTAR refers to the corrective optical system installed in
the HST during the 1993 December Space Shuttle repair mission. 
We observed the QSO in the 
FP-SPLIT= 4 sequence to reduce
the effects of detector/photocathode fixed pattern noise.  Detector
scanning step pattern number 4 provided two samples per diode width and
sampling of the object spectrum and the detector background with the full
500 channel array of the GHRS Digicon detector. We spent 
11\% of the observing time
on the background observations. For technical descriptions of
the GHRS, see the Space Telescope Science Institute Instrument Handbooks
(Solderblom \etal 1994).

We processed the data with the HST CALHRS  reduction system developed by
the GHRS science team.  The data reduction converts raw counts into count
rates, adjusting for pulse counting dead time losses, particle radiation,
dark count events and diode to diode sensitivity variations.   Because of
the low signal level in the individual spectra, we used a standard merge to
combine the fifty-six integrations into a single spectrum.   The line
spread function (LSF) for these SSA observations should be well represented
by a Gaussian with FWHM = 14 \kms~(Gilliland \etal 1992; Robinson \etal
1998), which is consistent with the profile width we measured for
wavelength calibration arc lamp observations
obtained before and after the object integrations.

In this analysis, we use the wavelength scale
given by WAVECAL exposures which we obtained during the GHRS observations of
QSO 1718+4807, and we use the reduction package written by the GHRS team.
We find a wavelength offset of 16.4 \kms~from the spectrum analyzed by
WCLFLV, the difference most likely arising in an offset between 
the default GHRS wavelength scale and the wavelength calibrations taken
between science exposures.  The WAVECAL calibrations gives a more accurate
scale for the GHRS spectrum and we use this scale throughout the paper.
The spectrum is shown in Figure 1,
along with a cubic spline which we use to fit the quasar continuum.

At the redshift of interest, $z = 0.7011$, two absorption lines
are seen in the GHRS spectrum: \Lya and \SiIII.
The diodes are 0.098 \AA~wide (or 14.2 \kms) and two samples
were taken per diode width. The signal-to-noise ratio (SNR) is 
approximately 9 per diode sample.
The \SiIII ($\lambda1206$) line is best fit with
a single Voigt profile at $z=0.701117 \pm 0.000008$ with a column
density, log N(\SiIII) = $12.85 \pm 0.05$ cm$^{-2}$, and velocity
dispersion, $b = 17 \pm 2$ \kms~($b = \sqrt{2} \sigma$).    
The column density and velocity dispersion for \SiIII measured
in our GHRS spectrum agree with WCLFLV, but the measured redshift 
is offset by 16.4 \kms~due to the different wavelength calibrations.

We obtained optical spectra of
QSO 1718+4807 with the HIRES spectrograph (\cite{vog94})
on the 10-m Keck-1 telescope, with spectral resolution of 8 \kms~(FWHM) and
full wavelength coverage from 3800 \AA~to 5000 \AA.
We performed a standard spectral extraction,
described by Tytler, Fan \& Burles (1996).
Three exposures totalling 4500 seconds
gave SNR = 50 per 2 \kms~pixel at the wavelengths of \MgII. As is 
common with most QSO metal-line systems, we detected 
\MgII~($\lambda\lambda$2796, 2803), but not 
\FeII~($\lambda\lambda\lambda$2344, 2382, 2600).
We measure log N(\MgII) = $11.5 \pm 0.1$ cm$^{-2}$, 
$z$(\MgII) = $0.701088 \pm 0.000007$, $b$(\MgII) = $13.5 
\pm 1.9$ \kms, and place an upper limit for the \FeII~
column density of log N(\FeII) $< 12.6$ cm$^{-2}$.  

Figure 2 shows the useful absorption lines at redshift $z=0.7011$.
The three metal lines, \SiIII~($\lambda$1206) 
and \MgII~($\lambda\lambda$2796,2803) are optically thin and best fit
with single Voigt profiles.  
The velocity difference between the \SiIII~and \MgII~lines is
$5.1 \pm 1.9$ \kms, where this $1\sigma$ error ignores errors in the
wavelength scales.
This difference is most likely due to the external error on the 
GHRS wavelength scale from the WAVECAL exposures, which has 
an RMS of $\Delta v = 3.5$ \kms~and a maximum
of $\Delta v = 10$ \kms, and hence the
\SiIII~and \MgII~may arise in the same gas.

The neutral hydrogen column density is well constrained by the
Lyman limit which has optical depth near unity (\cite{lan93}),
and is responsible for the continuous absorption below
1550 \AA~in the IUE spectrum (Figure 3).
We estimate the column density and 1$\sigma$ errors
log N(\HI) = $17.12 \pm 0.05$ cm$^{-2}$.  
The error in N(\HI) is much smaller than the final error in D/H.
Another hydrogen system at $z=0.602$ also contributes to the optical 
depth below 1460 \AA, with log N(\HI) $\simeq 16.7$ \cm2.
WCLFLV fit the quasar continuum below 1550 \AA~with 
just the system at $z = 0.701$, and found 
log N(\HI) = $17.24 \pm 0.01$ \cm2.  We suggest this column density
is too high,
because the lower redshift Lyman limit system accounts for some
of the absorption below 1460 \AA. 

We can estimate the temperature, metallicity, and 
neutral fraction in this absorption system.  We use velocity information
available for \HI~and \MgII~to convert equivalent widths of other ions into
approximate column densities. The velocity
dispersions are well constrained for \HI~and \MgII~by fitting single
Voigt profiles. For these two low-ionization species the observed columns
may, but need not, come from the same gas.
If we assume the temperature and turbulent velocity is identical for
both species, we find $T = 3.1 \pm 0.5 \times 10^4$ K and
$b_{tur} = 12.7 \pm 2.0$ \kms.

In archived spectra obtained with the HST Faint Object Spectrograph 
(R $\approx$ 1300 -- too small to give column densities), we measure 
equivalent widths $W_{obs}$(\CIV) = $0.13, 0.09$ \AA for the
\CIV~($\lambda\lambda1548,1550$) doublet at $z=0.701$.
Using the above values for $T$ and $b_{tur}$, this implies
a column density log N(\CIV) = $13.3 \pm 0.2$ cm$^{-2}$.
We place an upper limit on the column density of \ion{Si}{2}~
using a $2\sigma$ upper limit on
the equivalent width of \ion{Si}{2}~($\lambda1260$),
log N(\ion{Si}{2}) $<$ 12.3 cm$^{-2}$. 
We estimated the metal abundance using the photoionization code
CLOUDY (Ferland 1993). We used an ionizing background spectrum at $z = 0.7$
from Haardt \& Madau (1997), and
we assumed that the ratio of the two alpha elements Mg and Si is solar.
We find unremarkable parameters: an ionization parameter
log U $= -2.5$, a total hydrogen density $n_H = 10^{-3.1}$ ~cm$^{-2}$, 
and a neutral fraction of $n$(\HI)$/n_{\rm{H}} = 10^{-2.9}$, and a low
metal abundance [Si/H] $= -2.4$.

\section{Constraining D/H}

We measure D/H by comparing composite Voigt profile models with the QSO spectra.
We consider four models which differ in the number of Voigt profiles 
used, and other constraints. 
In each model, we adjust the parameters to minimize the $\chi^2$.  
The value of D/H depends varies a lot depending on the redshift(s) assumed
for the H. This should be measured from the high order Lyman series lines, 
which will be much less saturated than \Lya, and hence will hence show
more details of the velocity distribution. 
Since this data has not been obtained, we use the 
redshifts of the metal lines as constraints in Models 1 and 2.
In Table 1, we list the assumed redshift
in each model.  All models except 4 have a single velocity
component, which produces a blend of \HI~and \DI~lines.
In Model 3 we leave the redshift of the H free.
In Model 4, we allow H at a second redshift which
absorbs flux near $z$(\DI).
We do not use metal lines to constrain the velocity dispersion and temperature
of the H and D because this would be an unnecessary constraint, which could
be inaccurate, e.g. if variations in the ionization and metallicity 
give different parameters for the H and metals.

For each model we 
step through values of D/H, $-5.0 < \rm{log} (D/H) < -3.0$
and fit to the GHRS data between 
2066 \AA~$< \lambda <$ 2070 \AA, with a constraint on N(\HI) 
measured with the Lyman limit.
We perform a least squares minimization of all free parameters in the models
over a range of D/H (Burles \& Tytler 1998).
This gives a measure of $\chi^2$ as a function of the single
parameter D/H.  We calculate the most likely D/H and confidence levels
from the $\chi^2$ functions (Figure 4).
In table 1, we list the minimum $\chi^2_{\rm{min}}$, 
the corresponding value of D/H with 95\% confidence levels,
and the $\chi^2$ probability that the data came from the model.
		   
In Table 2 we show that other uncertainties in the measurement 
are much smaller than the 
differences in the models and the random errors, with the exception
of the one-sided uncertainty due to contaminating absorption.
We estimate the uncertainty as
$\Delta \rm{log} \left({{D} \over {H}}\right) = 
-\rm{log} \left(1 - {{N(C)} \over {N(D~I)}}\right) \rm{ln} (1 - P),$
where P is the probability of contamination by an absorber with
column density N(C).  With the above estimates from \MgII~systems,
this uncertainty can be as high as $\Delta \rm{log} (D/H) \approx 1$.

The unabsorbed intrinsic quasar continuum level is well determined:
an error of 5\% would give $\Delta \, \rm{log D/H} < 0.04$ in all models.
The \HI~and \DI~lines have
FWHMs greater than 41 \kms~and 36 \kms~respectively and the D/H is not
sensitive to changes in the line spread function, which calibration
spectra show is a Gaussian with FWHM = 14 \kms. 

\subsection{Is there a Second Component to H?}

There are two possible origins for the 
second hydrogen component: the random \Lya forest and
gas associated with the D/H system. We discuss the likelihood of each.
The low column density \Lya forest has not yet been measured
at these redshifts, but
we can estimate the line density directly from the GHRS spectrum in
Figure 1.  Assuming that the four unidentified lines are \HI~\Lya lines,
we estimate a line density ${{dN} \over {dz}} \simeq 166$ for \Lya
lines with column densities 12.5 \cm2~$<$ Log N(\HI) $<$ 14.5 \cm2~for $z = 0.7$.
We calculate the probability of one or more \Lya lines to fall within 20 
\kms~of the expected \DI~position, P $\simeq$ 1\%.
% This is the combination of two Poisson distributions
% One has the mean number of Lya lines = 4
% The other has the number of lines to get the mean within 20 km/s
% The combination of both is almost identical to just setting
% the first to a delta function at 4 lines. 
This probability is calculated after the inspection of the data
({\it a posteriori}). Chance events often give much lower values for
such probabilities.

We know less about the velocity structure of low column,
low redshift Lyman limit systems.
% HST Key project was only sensitive to high column lines (Bachall \etal 1994).
% And could say nothing about velocity structures less than 200 \kms.
We would like to know, given that an LLS has been seen, what is the 
probability that it has only one velocity component in \HI?
There is no relevant data on \HI, but Keck telescope HIRES spectra 
provide clues. Charlton \& Churchill (1998) 
found that all \MgII~systems in a sample
of 26 at low redshift ($z < 1$) show 2 or more components in \MgII.  
This sample was restricted to \MgII~lines with W$_r(2796) > 0.3$ \AA~and
would not have contained our \MgII~line, which is unusually weak with
W$_r(2796) = 0.013$ ~\AA.

A second sample of \MgII~lines with $0.02 < $ W$_r < 0.3$ \AA~was presented by
Churchill \etal (1998).
The redshift range is appropriate: 0.4 -- 1.2, peaking near 0.7.
About 40\% of these 30 systems have more than one component or show wide lines
indicative of velocity structure.
There are three reasons why the relevant fraction is $>40$ \%.
1. Many of their spectra are of low SNR
and would not show weak \MgII~components.
2. \HI~is often seen in gas where \MgII~is not detected in even the highest
current SNR spectra.
3. From the number of lines per unit redshift, they calculate that
74\% of their systems do not have enough \HI~to show Lyman breaks, and
hence they have less \HI~than the absorber in 1718+4807.
Clustering increases with \HI~column (Cristiani \etal 1997), and hence
we expect that the sub-sample with larger N(\HI), comparable to the
1718 absorber, will show more components.
On the other hand, 1718 has extremely weak \MgII~lines and it
was selected because the LLS seemed unusually
steep in IUE spectra, which should exclude the more complex systems,
and decrease the probability that 1718 would show more than one 
component.
In summary, the probability that an absorber with \HI~column
like that in 1718+4807 shows 
more than one component is in the range 40 -- 100\%. 
The probability that
the absorber in 1718 has more than one component will be lower than this,
because we know its LLS is steep, the \MgII~lines are weak, and
they and the \SiIII~line are simple, but we do not use this information 
because it would give an {\it a posteriori} probability,
which would be hard to interpret.

\section{SUMMARY}

This is the first QSO absorption system to give limits on D/H 
at a moderate redshift. Compared to other systems at $z>2.5$, this
D/H limit is less reliable because the low signal to noise in the
\Lya line and the lack of spectra of other Lyman series lines allow
a variety of models which give a wide range of possible D/H.
The reduced contamination by the random \Lya forest is an advantage
which does not compensate for the other two disadvantages.
 
Four lines are detected in high resolution spectra.  
\SiIII~and \MgII~show simple lines
which can come from a simple model with one velocity component
which gives Voigt line profiles. In this case the asymmetry of the
\lya line must be explained by another ion. The wavelength of the
asymmetric absorption is a good match to the expected location of D.
Our three models of this situation use different assumptions about the
redshift of the gas and give 95\% confidence levels in the range
$8 \times 10^{-5} < \rm{D/H} < 57 \times 10^{-5}$.  

The asymmetric \Lya line can equally well be modeled with a more
sophisticated velocity distribution.  Levshakov \etal (1998) 
found models which give D/H $\simeq 4.4 \times 10^{-5}$ if there
are significant correlations in the velocity field.
We present a model with gas at two velocities which gives the
upper limit $\rm{D/H} < 50 \times 10^{-5}$.  
Absorbers with similar \HI~columns show complex velocity structure
in 40 -- 100\% of cases. The probability that this absorber is complex
will be lower because we know that the \SiIII~and \MgII~lines
appear simple, and the Lyman limit is steep.
The dominant uncertainty in the D/H comes from our ignorance of 
the \HI~velocity field. If it is simple 
then the uncertainty in the redshift of this component, and random errors
from the low signal to noise each give about a factor of two
uncertainty (95\%) in D/H. 

We find a good fit to the metal lines using
a solar Mg/Si abundance ratio, and [Si/H] = $-2.4 \pm 0.1$.
Even though the system is at a much lower redshift than previous
D/H observations, the low metallicity suggests minimal 
chemical processing. 

A high value of D/H ($> 10^{-4}$) gives a low cosmic baryon
density, $\Omega_b < 0.01 h^{-2}$, where $H_0 = 100 \, h \, \rm{km s}^{-1}$.
In standard big bang nucleosynthesis, a value for D/H fixes the
cosmological baryon to photon ratio and predicts the primordial abundances
of other elements. The high D/H is consistent with some estimates of
the primordial $^4$He abundance (Pagel \etal 1992, Olive \& Steigman 1995,
Olive \etal 1997),
but not others (Izotov and Thuan 1998).
The high D/H is not consistent with galactic chemical evolution 
(Tosi \etal 1998)
or with other independent measures of $\Omega_b$, from
the Lyman-$\alpha$ forest
(Bi \& Davidsen 1997, Rauch \etal 1997, Weinberg \etal 1997,
Zhang \etal 1998), galaxy clusters (Schramm 1998),
and early data on the power spectrum of the cosmic microwave background
(Lineweaver \& Barbosa 1998).

Two other QSOs give low
values of D/H = $3.4 \, \pm \, 0.3 \times 10^{-5}$ 
(Burles \& Tytler 1998a,b) which is consistent
with all other QSO data, but not with the high D/H. Low D/H implies a high
baryon density consistent with other measures.
Clearly, new measures of D/H in other QSOs are urgently needed.

\acknowledgements

This work was funded in part by a grant from the Space Telescope
Science Institute, GO-5488 and by G-NASA/NAG5-3237.
We are grateful to Steve Vogt, the PI of the Keck HIRES instrument,
and Theresa Chelminiak, Barbara Schaefer
who assisted us in obtaining the Keck spectra.
We thank the anonymous referee whose insight and support improved
the manuscript. 

%\clearpage

%****************** STANDARD TABLE *******************************
% for centering Table 1

\begin{table*}   
\begin{center}
\title{D/H Absorption Models}
\begin{tabular}{cccccccc}
\noalign{\vskip 5pt}
\tableline
Model & Redshifts & $\chi^2_{min}$ & D/H ($-2\sigma$)\tablenotemark{a} &
D/H($\chi^2_{min}$)\tablenotemark{a} & D/H ($+2\sigma$)\tablenotemark{a} &
par\tablenotemark{b} & P($\chi^2_{min}$)\tablenotemark{c} \cr
\tableline
 1  & $z_1=z$(\SiIII) & 44.8 & $-3.78$ & $-3.60$ & $-3.44$ & 4 & 0.37 \cr  
 2  & $z_1=z$(\MgII) & 51.2 & $-4.08$ & $-3.86$ & $-3.62$ & 4 & 0.25 \cr  
 3  & $z_1$ free & 43.7 & $-3.80$ & $-3.46$ & $-3.24$ & 5 & 0.37 \cr  
 4  & $z_1, z_2$ both free & 42.0 & ...\tablenotemark{d} & ... & $-3.30$ & 8 & 0.30 \cr  

\tablenotetext{a}{In logarithmic units}
\tablenotetext{b}{Number of free parameters in model}
\tablenotetext{c}{Fit with 46 diode samples in GHRS spectrum covering \Lya}
\tablenotetext{d}{Lower limit extends to below parameter range of log(D/H) = -5.0}

\end{tabular}
\end{center}
 
\tablenum{1}
\end{table*}

%\clearpage

%****************** STANDARD TABLE *******************************
% for centering Table 2

\begin{table*}   
\begin{center}
\title{Uncertainties in the D/H measurement\tablenotemark{a} }
\begin{tabular}{cc}
\noalign{\vskip 5pt}
\tableline
Uncertainty & Magnitude\tablenotemark{b} \cr
\tableline
Choice of Model & $\pm 0.40$ \cr
Random Errors & $\pm 0.34$\tablenotemark{c} \cr
Total Hydrogen\tablenotemark{d} & $\pm 0.10$ \cr
Continuum Placement & $\pm 0.04$ \cr
Line Spread Function & $\pm 0.02$ \cr
Interloping Hydrogen &   $\approx -1.0 \times \rm{P_i}$\tablenotemark{e} \cr

\tablenotetext{a}{Corresponding to 95\% confidence}
\tablenotetext{b}{In logarithmic units of log D/H}
\tablenotetext{c}{Largest 95\% uncertainty error in Table 1}
\tablenotetext{d}{Estimated from Lyman continuum absorption}
\tablenotetext{e}{Depends on the probability of interloper, $\rm{P_i}$}

\end{tabular}
\end{center}
 
\tablenum{2}
\end{table*}

%\clearpage

%\section{Figure Captions}

\begin{figure}
\centerline{
\psfig{file=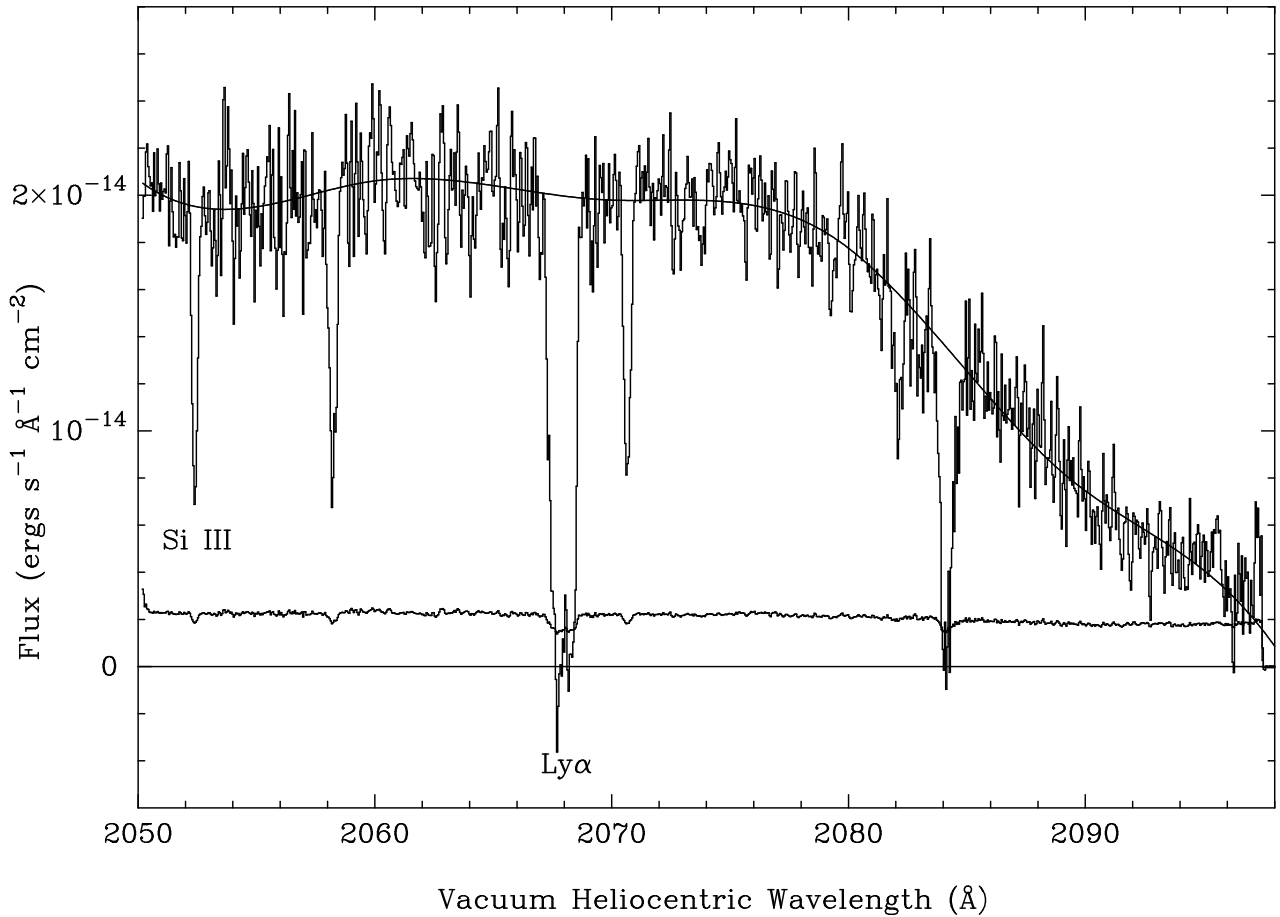,height=6.0in,angle=-90}}
\caption{
GHRS spectrum of QSO 1718+4807.  
The histogram shows the observed flux, the
lower line is the 1$\sigma$ error per diode sample, and the solid line is the
continuum fit.  The spectrum was obtained with 14.2 \kms~wide diodes with
two samples per diode width.
The \Lya feature of interest is seen at 2068 \AA, and the corresponding
\SiIII~line is at 2052 \AA.  The SNR is 9 per diode sample at \Lya and \SiIII.}
\end{figure}

\begin{figure}
\centerline{
\psfig{file=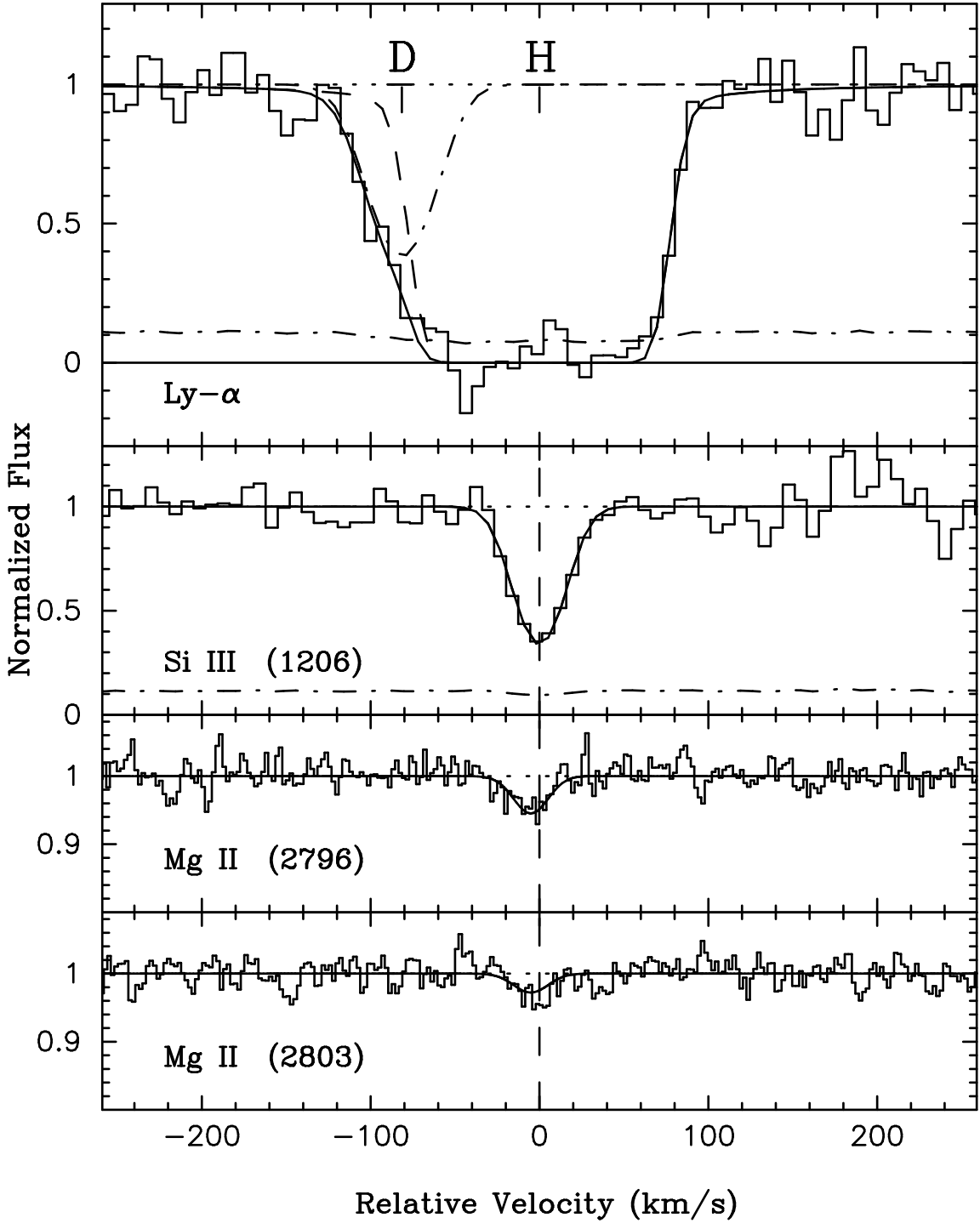,width=\columnwidth}}
\caption{
Velocity plots of Lyman $\alpha$, \SiIII~(1206), and
\MgII~(2796,2803) absorption features 
in the system towards QSO 1718+4807.
Zero velocity corresponds to the redshift $z = 0.701117$ measured for
\SiIII. 
The histogram represents the observed counts in each diode sample
normalized to the quasar continuum.  The smooth line shows a typical composite 
Voigt profile fit to the data.  }
\end{figure}

\begin{figure}
\centerline{
\psfig{file=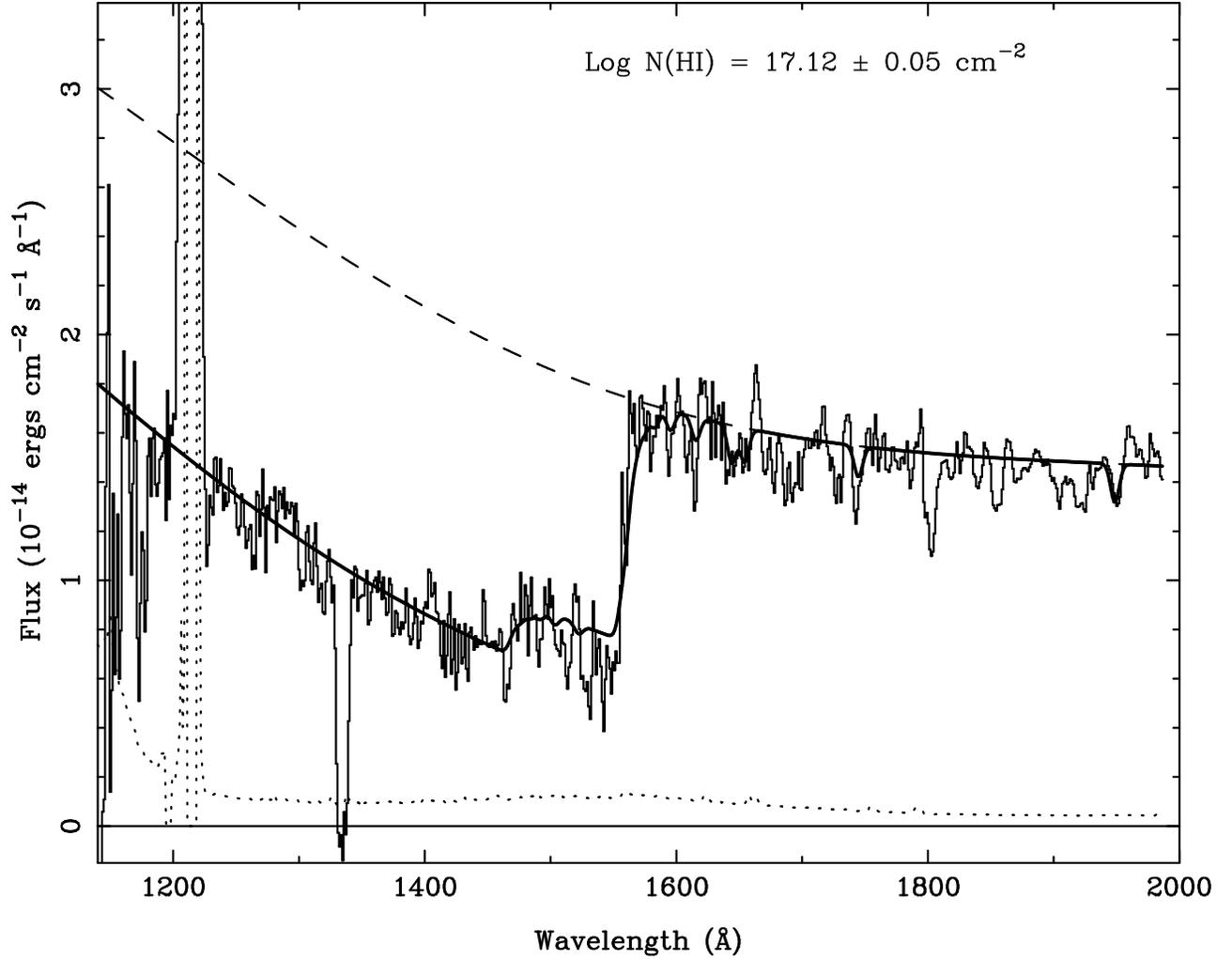,height=6.0in,angle=-90}}
\caption{
IUE spectrum of QSO 1718+4807 ($z_{em}=1.084$, V=15.3).
The Lyman limit arising in the
system at $z=0.7011$ is clearly seen as a continuum break near 1550 \AA.
The solid line represents hydrogen absorption with N(\HI) = 17.12 at this
redshift and N(\HI) = 16.7 at $z=0.602$.  The dotted line shows the
1$\sigma$ error array, and the dashed line represents the assumed
quasar continuum.}
\end{figure}

\begin{figure}
\centerline{
\psfig{file=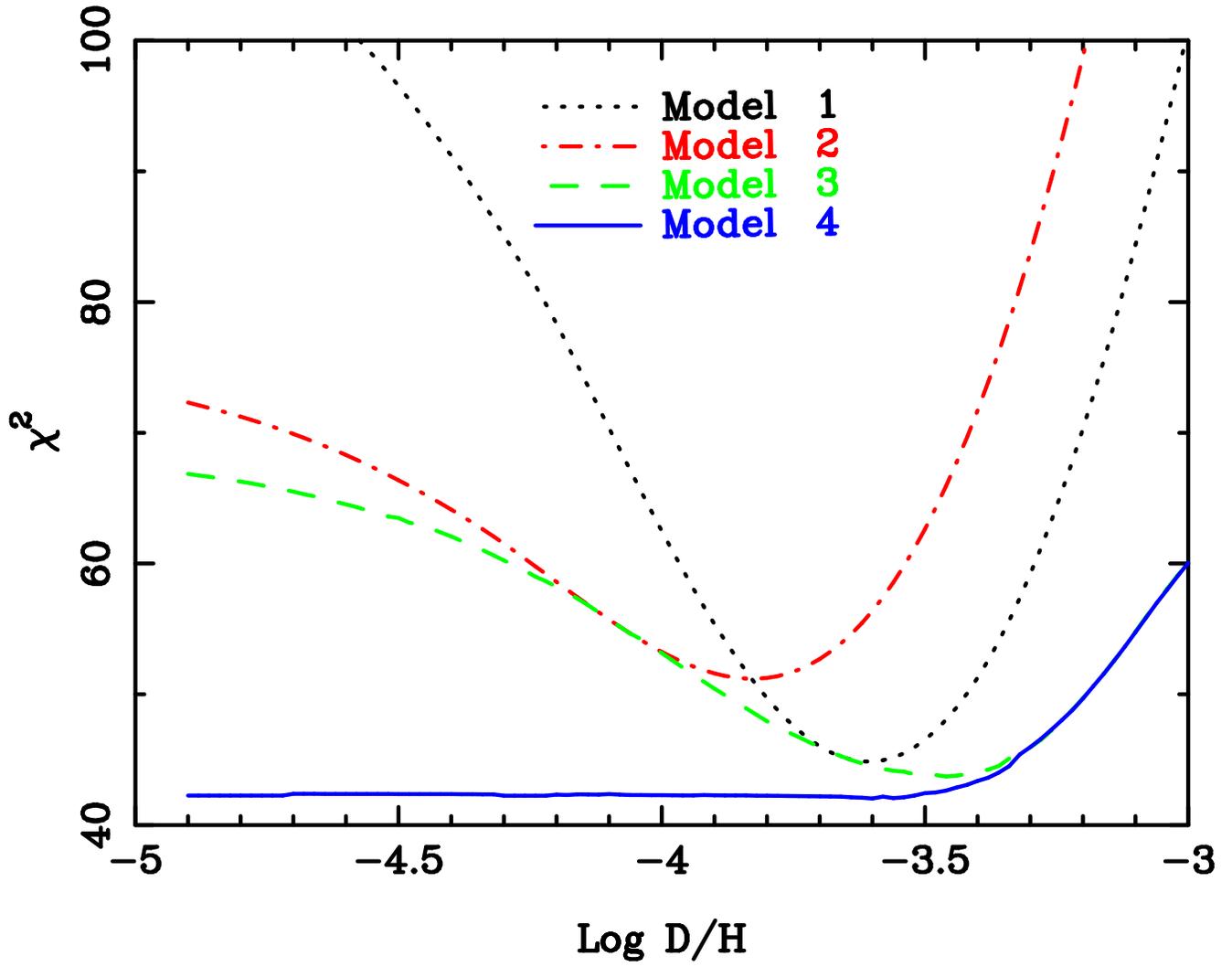,height=6.0in,angle=-90}}
\caption{ $\chi^2$ as a function of D/H for the 4 models.}
\end{figure}

\end{document}